 \documentclass[twocolumn,PRB,aps,psfig,preprintnumbers,superscriptaddress]{revtex4}
\usepackage{graphicx}
\usepackage{epstopdf}
\usepackage{float}
\usepackage{color}
\usepackage{amsmath}

\begin{document}

\title{Ferrimagnetism  in EuFe$_4$As$_{12}$ revealed by $^{153}$Eu NMR and $^{75}$As NQR measurements}
\author{Q.-P. Ding}
\affiliation{Ames Laboratory, U.S. Department of Energy, Ames, Iowa 50011, USA}
\affiliation{Department of Physics and Astronomy, Iowa State University, Ames, Iowa 50011, USA}
\author{H.-C. Lee}
\affiliation{Ames Laboratory, U.S. Department of Energy, Ames, Iowa 50011, USA}
\affiliation{Department of Physics and Astronomy, Iowa State University, Ames, Iowa 50011, USA}
\author{K. Nishine}
\affiliation{Muroran Institute of Technology, Muroran, Hokkaido 050-8585, Japan}
\author{Y. Kawamura}
\affiliation{Muroran Institute of Technology, Muroran, Hokkaido 050-8585, Japan}
\author{C. Sekine}
\affiliation{Muroran Institute of Technology, Muroran, Hokkaido 050-8585, Japan}
\author{Y. Furukawa}
\affiliation{Ames Laboratory, U.S. Department of Energy, Ames, Iowa 50011, USA}
\affiliation{Department of Physics and Astronomy, Iowa State University, Ames, Iowa 50011, USA}

\date{\today}

\begin{abstract} 
     Filled skutterudite compound EuFe$_4$As$_{12}$ shows the highest magnetic ordering temperature of $T_{\rm C}$ = 154 K among Eu-based skutterudite compounds,  but its magnetic ground state has not been determined yet. 
     Here, we performed  $^{153}$Eu nuclear magnetic resonance (NMR) and $^{75}$As nuclear quadrupole resonance (NQR) measurements on EuFe$_4$As$_{12}$ to reveal its magnetic ground state as well as the physical properties from a microscopic point of view.
     From the temperature and magnetic field dependence of $^{153}$Eu NMR spectrum in the magnetically ordered state, we found that the Eu ions are in Eu$^{2+}$ state with a nearly 7 $\mu_{\rm B}$ corresponding to $S$ = 7/2 spins. 
     Combined with the magnetization measurements which show the reduced saturation moments of 4.5 $\mu_{\rm B}$/f.u., we determined the ground magnetic structure in EuFe$_4$As$_{12}$ to be ferrimagnetic where the Eu$^{2+}$ 4$f$ and the Fe 3$d$ ordered moments are ferromagnetically aligned in each sublattice but the moments between the sublattices are antiferromagnetically aligned. 
   We also found the local distortion at the Eu site from the cubic symmetry in the magnetically ordered state.
     The relationship between the rattling motion of Eu atoms and the local symmetry of the Eu ions is discussed.
     From the $^{75}$As NQR nuclear spin-lattice relaxation time measurements as well as $^{153}$Eu NMR measurements, we found that the 4$f$ electrons of the Eu ions are well described by the local moment picture in both the magnetic and paramagnetic metallic states. 
     

\end{abstract}

\maketitle

 \section{Introduction} 

   The interplay of 4$f$ and itinerant electrons in rare-earth containing metallic systems provides fascinating physical phenomena such as superconductivity, ferromagnetism, antiferromagnetism, non-Fermi-liquid behavior, heavy-fermion states, and so on  \cite{Harima_Review,Stewart1984,Radousky,Muller2001, Sales2005,Kirchner2020}.
    Among many materials bearing rare-earth elements,  the europium containing Fe-based superconducting material EuFe$_2$As$_2$ stands as one of the interesting materials since, with carrier doping or pressure application, the system shows the coexistence of superconductivity and magnetism originating from Eu 4$f$ moments, where the magnetism of itinerant Fe moments also plays an important role \cite{Jeevan2008,Tereshima2009,Jin2016,Zapf2017}.  
    EuFe$_2$As$_2$ at ambient pressure exhibits two distinct magnetic phase transitions at $T$ = 189 K and 19 K. 
    The first magnetic order is a spin density wave [stripe-type antiferromagnetic (AFM) state] associated with the itinerant Fe moments while the second one is due to  Eu$^{2+}$ 4$f$ moments, making an A-type AFM structure where the Eu ordered moments are ferromagnetically aligned in the $ab$ plane but the moments in adjacent layers along the $c$ axis are antiferrmagnetically aligned \cite{Jeevan2008}.
     Although the Eu and Fe moments order at different temperatures, the strong coupling between the 4$f$ and itinerant electrons has been pointed out, which will be responsible for the interesting and complicated magnetic properties observed in Eu(Fe$_{1-x}$Co$_x$)$_2$As$_2$ where the A-type magnetic structure changes to the A-type canted AFM structure at intermediate Co doping levels around $x \sim$ 0.1, and then  to the ferromagnetic order along the $c$ axis at $x$ $\sim$ 0.18 \cite{Jin2016}. 
   
   The importance of the magnetic interaction between Eu 4$f$ electrons and itinerant $d$ electrons has also been pointed out in an Eu-based filled skutterudite compound EuFe$_4$As$_{12}$ \cite{Sekine2009}.
   However, the magnetism of EuFe$_4$As$_{12}$ can be quite different from the aforementioned Eu(Fe$_{1-x}$Co$_x$)$_2$As$_2$.
   EuFe$_4$A$_{12}$ exhibits a magnetic phase transition at $T_{\rm C}$ $\sim$ 152 K where both Eu 4$f$ and itinerant Fe moments order at the same time \cite{Sekine2009, Sekine2015}.
     It is pointed out that the transition temperature of $T_{\rm C}$ $\sim$ 152 K is relatively high in comparison with other related compounds  \cite{Sekine2009}. 
   When Fe is replaced by Ru or Os, the magnetic ordering temperature is suppressed to 25 K for  EuOs$_4$As$_{12}$ and no magnetic order is observed down to 2 K in EuRu$_4$As$_{12}$ \cite{Sekine2009}.
    On the other hand, when Eu$^{2+}$ is replaced by other divalent Sr or Ba ions, no magnetic order is observed although ferromagnetic spin fluctuations were reported \cite{Nishine2017,Sekine2015Sr, Ding2018}. 
    In the case of La for the replacement,  an itinerant ferromagnetism associated with Fe $3d$ moments is observed below a Curie temperature of $\sim$ 5.2 K in LaFe$_4$As$_{12}$ \cite{Tatsuoka2008, Namiki2010} .  
   The difference of the magnetic ordering temperature for those compounds indicates the strong exchange coupling between Eu 4$f$ electrons and itinerant Fe 3$d$ electrons \cite{Sekine2009}.
     Such strong exchange couplings have also been reported in similar Eu-containing iron skutterudite compounds  EuFe$_4$Sb$_{12}$ ($T_{\rm C}$ = 85 $\pm$ 4 K) \cite{Bauer2001,Takeda2000,Kihou2004,Danebrock1996,Bauer2004} and EuFe$_4$P$_{12}$ ($T_{\rm C}$ = 100 $\pm$ 3 K) \cite{Gerard1983,Grandjean1984}.  
     
     As for the spin structure of the magnetic state, magnetic susceptibility $\chi (T)$ measurements on EuFe$_4$As$_{12}$ suggest either a canted ferromagnetic or ferrimagnetic structure below $T_{\rm C}$ \cite{Sekine2009}. 
       The effective moment and Curie-Weiss temperature in the paramagnetic state  estimated from  $\chi (T)$ measurement are reported to be 6.93 $\mu_{\rm B}$/f.u. and 46 K, respectively.   
The  value of the effective moments is slightly smaller than that expected for divalent Eu$^{2+}$ ($S$ = 7/2) of $\mu_{\rm eff}$ = 7.94 $\mu_{\rm B}$ \cite{Sekine2009}.
   The magnetic field dependence of magnetization at 2 K for EuFe$_4$As$_{12}$ shows a saturation moment of 4.5 $\mu_{\rm B}$/f.u. at 1 T, which is much smaller than 7 $\mu_{\rm B}$ expected for Eu$^{2+}$ \cite{Sekine2009}.
     Similar magnetic properties have been reported in the isostructural compound EuFe$_4$Sb$_{12}$ which was initially considered as a ferromagnet below $T_{\rm C}$ \cite{Bauer2001,Takeda2000,Danebrock1996}.  
     Later on, the x-ray magnetic circular dichroism spectroscopy (XMCD) and x-ray absorption spectroscopy (XAS) measurements suggest a ferrimagnetic state in EuFe$_4$Sb$_{12}$ where ferromagnetically aligned Eu spins are ordered antiferromagnetically with ordered Fe moments \cite{Krishnamurthy2007},  however, the possibility of a canted  ferromagnetic state was not fully ruled out. 
   In the case of EuFe$_4$A$_{12}$, such measurements have not been performed yet. 
     Therefore, the detailed studies of the local magnetic and electronic properties of the magnetic ions are important to reveal the magnetic structure as well as the magnetic properties of the system.

    Nuclear magnetic resonance (NMR) is a powerful technique to investigate magnetic properties of materials from a microscopic point of view. 
   In particular, one can obtain direct and local information of magnetic state at nuclear sites, helping to determine the magnetic structure of magnetic systems. 
   Although the magnetic state of the Eu ions in EuFe$_4$As$_{12}$ is a key to understand the magnetic properties of the system, there have been no Eu NMR studies of this compound up to now to our knowledge.

    In this paper, we have carried out $^{153}$Eu NMR and $^{75}$As nuclear quadrupole resonance (NQR)  measurements  to investigate the magnetic properties of EuFe$_4$As$_{12}$, especially focusing on the magnetic state of the Eu ions,  from a microscopic point of view.
      From the external magnetic field dependence of $^{153}$Eu NMR spectrum, the Eu ions are shown to be in divalent state with nearly 7 $\mu_{\rm B}$  which ordered ferromagnetically without any canting component, revealing a ferrimagnetic ordered state in EuFe$_4$As$_{12}$.
     We also report the local distortion at the Eu site from the cubic symmetry in the magnetically ordered state, which will be important to understand the physics of motion for Eu ions called rattling in the cage formed by Fe$_4$As$_{12}$ units.

 \section{Experimental}
 
    Polycrystalline EuFe$_4$As$_{12}$ samples  were prepared at high temperature and high pressures using a wedge-type cubic anvil high-pressure apparatus \cite {Sekine2009}. 
        The chemical composition of the samples was determined  by energy-dispersive x-ray analysis (EDX) using a JEOL  scanning electron microscope and was  found to be  Eu$_{1.06}$Fe$_{4.00}$As$_{12.2}$ showing no obvious deficiency of Eu ions.  Good chemical homogeneity was found in the powder samples. The crystal structure of the sample was characterized by powder x-ray diffraction (XRD) using Mo $K_\alpha$ radiation and silicon as a standard.
   The powder samples are loosely packed into a sample case, which allows the crystallites to be oriented along the applied magnetic field direction.   
    NQR measurements of $^{75}$As ($I$ = $\frac{3}{2}$, $\frac{\gamma_{\rm N}}{2\pi}$ = 7.2919 MHz/T, $Q=$ 0.29 barns) and NMR measurements of $^{153}$Eu ($I$ = $\frac{5}{2}$, $\frac{\gamma_{\rm N}}{2\pi}$ = 4.632 MHz/T, $Q=$ 2.49 barns) nuclei were conducted using a homemade phase-coherent spin-echo pulse spectrometer. 
    $^{153}$Eu NMR spectra in zero and nonzero magnetic fields $H$ in the magnetically ordered state and $^{75}$As-NQR spectra were measured in steps of frequency $f$ by measuring the intensity of the Hahn spin echo. 
   The $^{75}$As and $^{153}$Eu nuclear spin-lattice relaxation rate 1/$T_ 1$ was measured with a saturation recovery method.
   $1/T_1$ at each temperature was determined by fitting the nuclear magnetization $M$ versus time $t$  using the exponential function $1-M(t)/M(\infty) = e^ {-(3t/T_{1})}$ for $^{75}$As NQR, and $1-M(t)/M(\infty) = 0.029e^ {-(t/T_{1})}+0.18e^ {-(6t/T_{1})}+0.79e^ {-(15t/T_{1})}$ for $^{153}$Eu NMR, where $M(t)$ and $M(\infty)$ are the nuclear magnetization at time $t$ after the saturation and the equilibrium nuclear magnetization at $t$ $\rightarrow$ $\infty$, respectively.

  \section{Results and discussion}
\subsection{$^{153}$Eu zero field NMR in the Ferrimagnetic state}

\begin{figure}[b]
\includegraphics[width=\columnwidth]{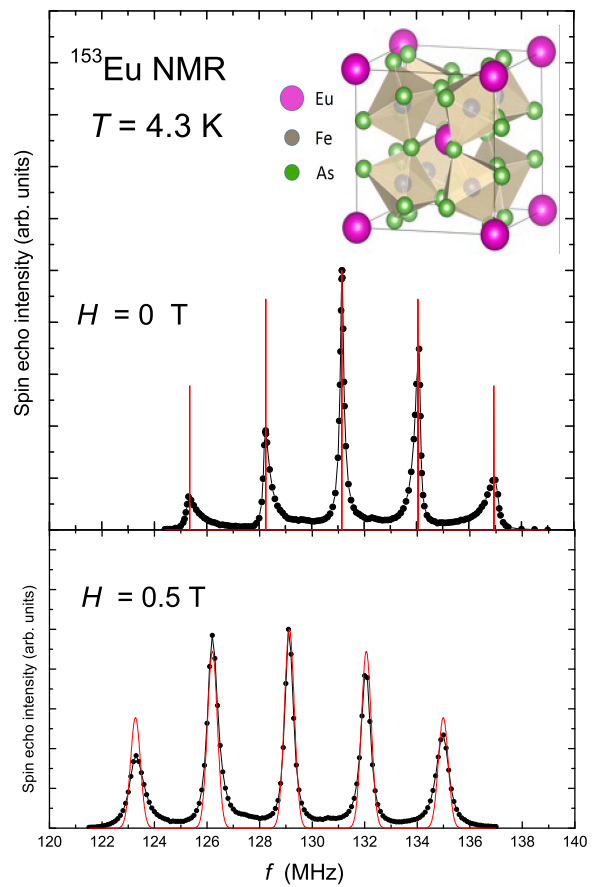} 
\caption{Top:  $^{153}$Eu-NMR spectrum in zero magnetic field at $T$ = 4.3 K in the magnetic ordering state. 
  The lines are the calculated positions of $^{153}$Eu-NMR spectrum using $B_{\rm int}$ = -28.14 T and $\nu_{\rm Q}$ = 2.90 MHz. 
The inset shows the crystal structure. 
     Bottom: $^{153}$Eu-NMR spectrum under $H$ = 0.5 T. The curves are calculated spectrum using the same parameters  where a line broadening of 0.35 MHz is introduced for each line. 
     }
\label{fig:EuNMR}
\end{figure}

   Figure  \ \ref{fig:EuNMR} shows the frequency-swept $^{153}$Eu NMR spectrum in zero magnetic field at 4.3 K in the magnetically ordered state. 
  The observation of the $^{153}$Eu NMR signals clearly evidences that the magnetic moments of Eu 4$f$ electrons order in the magnetic state. 
   The relatively sharp peaks in the observed spectrum indicate a high quality of the powder samples. 
    The peak positions of the spectrum are well explained by the combination of a large Zeeman interaction due to magnetic field [for the present case, an internal magnetic induction ($B_{\rm int}$) at the Eu site] and a small quadrupole interaction whose nuclear spin Hamiltonian is given as follows;
\begin{eqnarray}
\centering
{\cal H} &=& -\gamma_{\rm n}\hbar  {\bf I} \cdot {\bf B_{\rm int}}  + \frac{h\nu_{\rm Q}}{6}[3I_Z^2-I^2 + \frac{1}{2}\eta(I_+^2 +I_-^2)],
\label{eq:1}
\end{eqnarray} 
where
\begin{equation}
I_z = \frac{1}{2}(I_+ e^{-i\phi}+I_- e^{i\phi}){\rm sin}\theta+I_Z{\rm cos}\theta.
\label{eq:quantization}
\end{equation}
Here $h$ is Planck's constant, and $\nu_{\rm Q}$ is nuclear quadrupole frequency defined by $\nu_{\rm Q} = 3e^2QV_{ZZ}/2I(2I-1)h$  $(=3e^2QV_{ZZ}/20h$ for $I$ = 5/2)  where $Q$ is the electric quadrupole moment of the Eu nucleus, $V_{ZZ}$ is the electric field gradient (EFG) at the Eu site  in the coordinate of the principal $X$, $Y$, and $Z$ axes of EFG, and $\eta$ is the asymmetry parameter of the EFG \cite{Slichterbook}.
   $\theta$ and $\phi$ are the polar and azimuthal angles between  the $Z$ axis of EFG and the direction of $B_{\rm int}$, respectively, where the quantization axis ($z$ axis) for the Zeeman interaction is pointing along the $B_{\rm int}$ direction. 
 

\begin{table*}
\caption{$B_{\rm int}$ and $\nu_{\rm Q}$ from $^{153}$Eu NMR measurements at 4.2 K for EuFe$_4$As$_{12}$, the helical antiferromagnets (AFM)  EuCo$_2$P$_2$ and EuCo$_2$As$_2$, and the  A-type antiferromagnet EuGa$_4$ and the magnetic moments of Eu ions $M_{\rm Eu}$ from neutron diffraction measurements on EuCo$_2$P$_2$ \cite{Reehuis1992} and EuCo$_2$As$_2$ \cite{Tan2016}.} 
\label{Table:Parameter}
 \begin{ruledtabular}
		\begin{tabular}{l c c c c c l }
		 &  $B_{\rm int}$ (T) &  $\nu_{\rm Q}$ (MHz)  &$M_{\rm Eu}$ ($\mu_{\rm B}$)  & Ground state  (ordered temperature)  & Ref.\\
		\hline
		 EuFe$_4$As$_{12}$ &  -28.14(5) T  & 2.90(5)  &    & Ferrimagnet  ($T_{\rm C}$ = 154 K) & This work \\
		 EuCo$_2$P$_2$    &  -27.5(1) T  & 30.6(1)  & 7.26  & Helical AFM ($T_{\rm N}$ =  45 K)  & \cite{Higa2017}\\
		 EuCo$_2$As$_2$  &  -25.75(2)T  & 30.2(2)  & 6.9(1)  & Helical AFM ($T_{\rm N}$ =  66.5 K)  & \cite{Ding2017}\\
		 EuGa$_4$     &  -27.08 T  & 30.5  &   & A-type AFM  ($T_{\rm N}$ =  16 K) & \cite{Yogi2013}\
	\end{tabular}
\end{ruledtabular}
\end{table*}

    In the case of $I$ = 5/2,  when $\eta$ = 0,  the NMR spectrum is composed of a central transition line  ($I_z$ = 1/2 $\leftrightarrow$ -1/2)  and  two pairs of satellite lines shifted from the central transition line by $\pm\frac{1}{2} \nu_{\rm Q}(3\cos^2\theta -1 )$ (for the transitions of $I_z$ = 3/2 $\leftrightarrow$ 1/2 and -3/2 $\leftrightarrow$ -1/2), and  $\pm \nu_{\rm Q}(3\cos^2\theta -1 )$ (for $I_z$ = 5/2 $\leftrightarrow$ 3/2 and -5/2 $\leftrightarrow$ -3/2).    
      The five solid lines shown in Fig.~\ref{fig:EuNMR} are the calculated positions using $|B_{\rm int}|$  = 28.14(5) T, $\nu_{\rm Q}$ = 2.90(5)  MHz, $\eta$ = 0  and $\theta = 0^\circ$ (and, thus, $\phi = 0^\circ$)  which reproduce the observed positions very well.
   Since $B_{\rm int}$ mainly originates from core polarization from 4$f$ electrons and is oriented in a direction opposite to that of the Eu spin moments, the sign of $B_{\rm int}$ is considered to be negative \cite{Freeman1965}.
   The observed $B_{\rm int}$ =  -28.14(5) T is close to the values of $B_{\rm int}$ reported from $^{153}$Eu zero-field NMR in the helical antiferromagnets EuCo$_2$P$_2$ \cite{Higa2017} and EuCo$_2$As$_2$ \cite{Ding2017}, and the A-type antiferromagnet EuGa$_4$ \cite{Yogi2013} as shown in Table 1.
    The values of Eu magnetic moments in EuCo$_2$P$_2$ and EuCo$_2$As$_2$ determined by neutron diffraction measurements are close to 7 $\mu_{\rm B}$ expected for Eu$^{2+}$ ion ($S$ = 7/2) \cite{Reehuis1992, Tan2016}. 
     Since the Eu ordered moment $M_{\rm Eu}$ is proportional to $|$$B_{\rm int}$/$A_{\rm hf}$$|$  where $A_{\rm hf}$ is the hyperfine coupling constant mainly originating from the core polarization,  the similar value of  $B_{\rm int}$ =  -28.14(5) T in EuFe$_4$As$_{12}$ in comparison with  those in other Eu compounds leads to the conclusion that the Eu ions are in Eu$^{2+}$ state with $S$ = 7/2.

    It is noted that, from the XRD measurements \cite {Sekine2009}, EuFe$_4$As$_{12}$ crystallizes in a body-centered cubic structure (the space group $Im\overline{3}$, see the inset of Fig. \ \ref{fig:EuNMR} \cite{VIETA}) with a lattice constant of 8.3374 \AA~at room temperature. 
   Since the local symmetry of the Eu site is cubic in the structure ($T_h$),  one expects no quadrupole interaction because the EFG due to the charges on the neighboring ions is zero.
   In general, there is another contribution to the EFG at the Eu site originating from the on-site $f$ electrons.
   However, this EFG contribution is also zero because of the spherical charge distribution of 4$f^7$ electrons for Eu$^{2+}$ ions with zero angular momentum $L$ = 0, which is independent of crystal structure.
   Therefore, the finite quadrupole interaction at the Eu site observed in the magnetically ordered state must be attributed to the contribution from the neighboring ions.
   Thus, the experimental data clearly evidence  the lowering symmetry at the Eu site from cubic.  
 We will discuss this issue later. 

       The direction of $B_{\rm int}$ from the NMR spectrum is not determined in the present case. 
  Usually, one can determine the direction from the value of $\theta$ if we know the principal axis of the EFG.
  However, as described above, our NMR spectrum indicates that the local symmetry at the Eu site in the magnetically ordered state is different from what expected from the high temperature cubic one.
   Therefore, the EFG direction cannot be determined, making the determination of the direction for $B_{\rm int}$ impossible at present. 

    It is also important to point out that the the satellite lines of the observed $^{153}$Eu zero-field NMR  are asymmetric and are tailing toward to the central transition line. 
    This asymmetric shape is reminiscent of a so-called powder pattern of NMR spectrum.   
      Although, in general, we do not expect the powder-pattern like shape in zero-field NMR spectrum, 
      we have the following three possibilities: the first one is due to the distribution of $\nu_{\rm Q}$, the second one comes from the slight distribution of $\theta$, and the last one originates from domain walls.
      The first scenario has been used to explain the similar asymmetric $^{153}$Eu zero-field NMR lines observed in EuGa$_4$ where a log-normal distribution of $\nu_{\rm Q}$ was used \cite{Yogi2013}. 
      However, we do not find such large distribution of $\nu_{\rm Q}$ in our case, as will be shown below. 
      As for the second scenario, as the small deviation in $\theta$ from 0 degree makes the satellite line shift toward the central transition line as described in the above equations for the satellite line positions, one may explain the characteristic asymmetric shape of the satellite lines by taking the small deviation from $\theta$ = 0$^{\circ}$ into consideration. 
      In fact, such scenario has been applied to explain the asymmetric lines observed in EuAl$_4$ where  the distribution of $\theta$ is estimated to be at most 6$^{\circ}$ or less \cite{Niki2020}.
      However, for our case, this seems to be not the case. 
      Since the signal intensity does not go down to zero between the satellite lines, to reproduce the spectrum, one needs to have a relatively large distribution of $\theta$ up to at least $\sim$ 20$^{\circ}$, which seems to be too large.
    In contrast, the third scenario can be the main reason for the observed asymmetric shape since the asymmetric shape of the lines becomes more symmetric when magnetic field is applied (a typical spectrum measured at $H$ = 0.5 T is shown at the bottom of Fig. \ \ref{fig:EuNMR}). 
     The red curves in the figure are the calculated spectrum at $H$ = 0.5 T with the same values of $B_{\rm int}$ and $\nu_{\rm Q}$ where we assume a line broadening of 0.35 MHz for each line.   
    It is noted that, in a magnetic field of 0.5 T, the values of line width (full width at half maximum,  $FWHM$ $\sim$ 0.4-0.5 MHz)  for each line are nearly the same, as the calculated spectrum reproduces the observed spectrum very well. 
    This indicates that the broadening of each line originates from the slight distribution of $B_{\rm int}$ (less than $\sim$ 0.4 \%)  and the effect of the distribution in $\nu_{\rm Q}$ into the line width is almost negligible.  
    It is known that the deviation of rare-earth filling factor from unity in filled skutteruride compounds produces the distribution of $\nu_{\rm Q}$ and thus broadening of spectrum at the rare-earth site \cite{Gippius2006, Yamamoto2008, Magishi2006}. 
    Therefore, the negligible small distribution of $\nu_{\rm Q}$ in EuFe$_4$As$_{12}$ suggests that the Eu filling factor is close to unity which is consistent with the results of the EDX measurements, confirming again the high quality of our samples.

\begin{figure}[tb]
\includegraphics[width=\columnwidth]{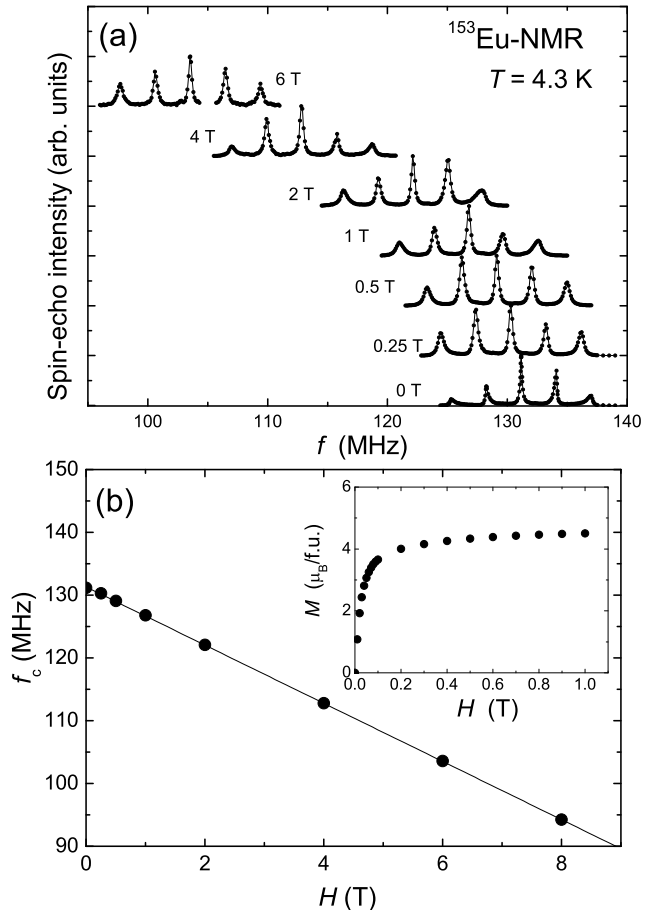} 
\caption{(a) Field dependence of $^{153}$Eu-NMR spectrum at 4.3 K. 
(b) Field dependence of $^{153}$Eu-NMR central line frequency ($f_{\rm c}$) at 4.3 K. 
The solid line is a linear fit whose slope is exactly same as -$\gamma /2\pi$ of $^{153}$Eu nucleus.
The inset shows the field dependence of magnetization measured at 2 K.
}
\label{fig:HEuNMR}
\end{figure} 

  \subsection{Magnetic field dependence of $^{153}$Eu NMR spectrum} 
    
     In order to gain more insight into the magnetic properties of EuFe$_4$As$_{12}$, especially, the Eu ordered moments in the magnetically ordered state, we have measured $^{153}$Eu NMR spectrum using the loosely packed sample under an external magnetic field $H$.
    Figure ~\ref{fig:HEuNMR}(a) shows the magnetic field dependence of the $^{153}$Eu NMR spectrum at 4.3 K. 
   The spectrum shifts to lower frequency as magnetic field increases. 
  At the same time, as discussed above, the asymmetric shape of the satellite lines become more symmetric under magnetic field, indicating the main reason for the asymmetry can be due to domain walls where the Eu ordered moments change the direction making the distribution of $\theta$. 
  It is also noted that the spacings between the lines keep constant, indicating $\theta$ does not change with the application of magnetic field. 
   Given the total magnetization saturates around 1 T at $T$ = 2 K as shown in the inset of Fig. ~\ref{fig:HEuNMR}(b), these results indicate that most of the small particles are aligned along the magnetic field direction. 
  This means there must be a finite magnetic anisotropy in the magnetic ordered state, although we cannot estimate the magnitude of it.
   
     The magnetic field  dependence of the resonance frequency for the central line ($f_{\rm c}$) is shown in Fig. ~\ref{fig:HEuNMR}(b) and the slope of the $H$ dependence of $f_{\rm c}$ is found to be -4.63 MHz/T which is exactly the same as -$\gamma /2\pi$ of $^{153}$Eu nucleus. 
    Since the effective field at the Eu site is given by the vector sum of $B_{\rm int}$ and $H$, i.e., $|B_{\rm eff}|$ = $|$$B_{\rm int} + H $$|$, the resonance frequency is expressed as $f$ = $\gamma /2\pi$ $|B_{\rm eff}|$.
    Therefore, the value of the slope clearly indicates that the direction of $B_{\rm int}$ is antiparallel to that of $H$ and also that all Eu ordered moments align along $H$ without any appreciable deviation.
  Thus, one can conclude that  the ferromagnetically aligned Eu ordered moments do not have any canting components up to 8 T, excluding clearly a canted ferromagntic state as a possible ground state.
   To explain the saturated magnetic moment of 4.5 $\mu_{\rm B}$/f.u. in the magnetic ordering state [see the inset of Fig. ~\ref{fig:HEuNMR}(b)], the Fe 3$d$ electrons must be magnetically ordered in antiparallel direction with respect to the Eu$^{2+}$ ordered moments, producing  the ferrimagnetic ground state. 
   Since the Eu ordered moments are estimated to be $\sim$ 7 $\mu_{\rm B}$ from the magnitude of $B_{\rm int}$ as discussed above, the Fe 3$d$ ordered moments are estimated to be $\sim$ 0.6 $\mu_{\rm B}$ for each Fe ion.
   A similar ferrimagnetic state has also been reported in the isostructural compound EuFe$_4$Sb$_{12}$ by XMCD and XAS measurements \cite{Krishnamurthy2007}.
   The ferrimagnetic state is consistent with the results from density function theory based calculations for EuFe$_4$As$_{12}$ \cite{ShanlerEuFe4As12} as well as other similar compounds such as EuFe$_4$P$_{12}$ \cite{ShanlerEuFe4P12} and EuFe$_4$Sb$_{12}$ \cite{ShanlerEuFe4Sb12}.


    \subsection{Temperature dependence of $^{153}$Eu zero-field NMR} 
       
\begin{figure}[tb]
\includegraphics[width=\columnwidth]{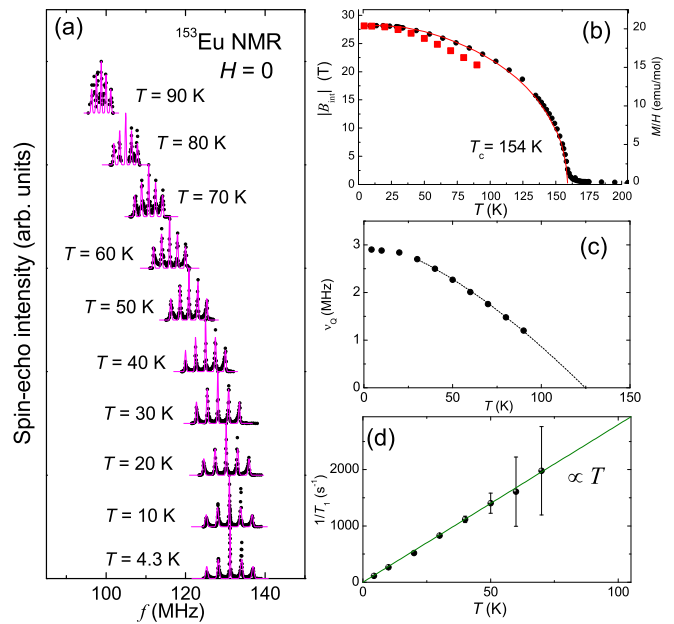} 
\caption{(a) Temperature dependence of $^{153}$Eu-NMR spectrum under zero magnetic field in the magnetic ordering state. 
  The curves are the calculated $^{153}$Eu-NMR spectra. 
     (b) Temperature dependence of $|B_{\rm int}|$. 
     The black circles show the temperature dependence of $M/H$ measured at $H$ = 0.1 T reported previously \cite{Sekine2009}.  The red curve is the Brillouin function with $J$ = $S$ = 7/2 and $T_{\rm C}$ = 154 K. 
     (c) Temperature dependence of $\nu_{\rm Q}$ at the $^{153}$Eu site. 
     (d) Temperature dependence of $^{153}$Eu 1/$T_1$ under zero magnetic field. The straight line shows the Korringa relation $1/T_1T$ = 28  (sK)$^{-1}$.
}
\label{fig:TEuNMR}
\end{figure}  
             
             Figure  \ \ref{fig:TEuNMR}(a)  shows the temperature dependence of $^{153}$Eu zero-field NMR spectra, where the spectra shift to lower frequency with increasing temperature.
   This is due to the reduction of $|B_{\rm int}|$ which decreases from 28.14 T at 4.3 K to 21.19 T at 90 K. 
   The red squares shown in Fig. ~\ref{fig:TEuNMR}(b) exhibit the temperature dependence of $|B_{\rm int}|$. 
   In the figure, we also plotted the $M/H$ data measured at 0.1 T by black circles \cite{Sekine2009}.
   As shown by the red curve, the temperature dependence of $M/H$ is well reproduced by the Brillouin function which was calculated based on the Weiss molecular field model with $J$ = $S$ = 7/2 and $T_{\rm C}$ = 154 K. 
   This suggests that, although both the Eu and Fe ordered moments contribute to the total magnetization, the temperature dependence of $M/H$ can be mainly characterized by the properties of Eu ordered moments. 
   In addition, the results indicate that the Eu ordered moments are well described by a local moment picture even in a metallic state as revealed by the resistivity measurements \cite{Sekine2009,Sekine2015} as well as nuclear relaxation measurements described below.
   As shown in Fig. ~\ref{fig:TEuNMR}(b), on the other hand, the temperature dependence  of $|B_{\rm int}|$ slightly deviates from that of $M/H$. 
   As $|B_{\rm int}|$ is proportional to $A_{\rm hf}$$M_{\rm Eu}$,  $|B_{\rm int}|$ should scale with $M/H$ since its temperature dependence is well explained by the $S$ =7/2 local moment picture. 
   Although the reason for the deviation is not clear at present, it may be due to the local distortion at the Eu site in the ferrimagnetic state. 
   As described above, the $|B_{\rm int}|$ is mainly determined by the hyperfine coupling originated from the 4$f$ electron core-polarization mechanism which is atomic in nature. 
   Therefore, the hyperfine coupling constant will not be affected by the local environment at the Eu site. 
   On the other hand, if one considers the effects of the transferred hyperfine field $B_{\rm int}^{\rm trans}$ at the Eu site from the nearest neighbor Fe ordered moments, the $B_{\rm int}^{\rm trans}$ can be affected by the local distortion at the Eu site since the transferred hyperfine field largely depends on the strength of Fe-As-Eu covalent bond.
  Although we cannot conclude the origin of the deviation, it would be interesting if it originates from the effects of $B_{\rm int}^{\rm trans}$ since this may be experimental evidence showing the coupling between the 3$d$ electrons of Fe and the 4$f$ electrons of Eu in EuFe$_4$As$_{12}$.

    The temperature dependence of $\nu_{\rm{Q}}$ at the $^{153}$Eu site is shown in Fig. \ \ref{fig:TEuNMR} (c) where  $\nu_{\rm{Q}}$ decreases from 2.90 MHz at 4.3 K to 1.2 MHz at 90 K. 
    Such a huge reduction of $\nu_{\rm{Q}}$ by $\sim$ 59 \% cannot be explained by thermal lattice expansion \cite{alphaQ} which is normally described by an empirical relation $\nu_{\rm Q}(T) = \nu_{\rm Q}(0)(1-\alpha_{\rm Q} T^{3/2})$ with a fitting parameter $\alpha_{\rm Q}$.
    As described above, one does not expect the finite value of $\nu_{\rm Q}$ at the Eu site in the high-temperature body-centered cubic structure. 
    In fact, $\nu_{\rm Q}$ seems to disappear around $T$ $\sim$ 125 K estimated from the smooth extrapolation using the experimental data at $T$ = 30 - 90 K, and $\nu_{\rm Q}$ will be zero in the paramagnetic state above $T_{\rm C}$ = 154 K where the cubic structure ($Im\overline{3}$) was determined by the XRD measurements at room temperature \cite{Sekine2009}.    

 Since the finite values of $\nu_{\rm Q}$ reflecting the degree of the local distortion of the cubic symmetry at the Eu site suggest a structural phase transition, we have carried out XRD measurements in the temperature range of $T$ = 113 - 300 K. 
 We did not observe clear evidence for structural phase transition in the XRD patterns down to 113 K which is the lowest temperature we can achieve by using our XRD spectrometer with a nitrogen gas flow type cryostat. 
   Although the results may suggest no structural phase transition at least down to 113 K, this does not fully rule out the possibility in the compound. 
   One of the possibilities is that the lowest temperature of 113 K may not be low enough to detect the structural phase transition which could occur at $T_{\rm C}$ estimated  from the temperature dependence of $\nu_{\rm Q}$.
   The situation here could be similar to the electron diffraction measurements of the skutterudite compound PrRu$_4$P$_{12}$ where the superlattice spots due to the structural phase transition were clearly observed at 12 K, however, the intensity of the superlattice spots becomes very weak at 40 K  even though the structural transition temperature is around 60 K \cite{PrRu4P12_0}. 
  Another possibility is that ordinary XRD measurements using powder samples may not be able to detect the structural phase transition.
  This has also been reported in PrRu$_4$P$_{12}$ where only synchrotron radiation XRD measurements using single crystals revealed a structural phase transition \cite{PrRu4P12_1,PrRu4P12_2}. 
   The structural phase transition was reported to change only in the space group (from $Im\overline{3}$ for the hight temperature phase to $Pm\overline{3}$ for the low temperature phase) while the cubic crystal symmetry of structure is unchanged \cite{PrRu4P12_0,PrRu4P12_1,PrRu4P12_2}. 
   It is interesting to point out that, even in the cubic symmetry, the slight displacements of P and Fe atoms have been reported in the low temperature phase of PrRu$_4$P$_{12}$ \cite{PrRu4P12_0,PrRu4P12_1,PrRu4P12_2}. 
   Such displacement may produce the local distortion of the cubic symmetry at the rare-earth sites, giving rise to the finite EFG. 
   This could explain our results in EuFe$_4$As$_{12}$, of course, we cannot conclude it though. 
   Thus, it is important to carry out the low temperature XRD measurements, especially using single crystals if available, to clarify whether or not the crystal structure in the magnetically ordered stat is different from the high temperature one. Such experiments are highly called for. 

    From the behavior of the temperature dependence of $\nu_{\rm Q}$, we consider that  the structural phase transition would a second-order type of phase transition if the local distortion at the Eu sites were due to a structural phase transition. 
      Since the specific heat measurements do not show any anomaly other than at $T_{\rm C}$ \cite{Sekine2015},  one may speculate that the magnetic ordering takes place with a possible concomitant structural phase transition, although the estimated $T$ $\sim$ 125 K is a little bit far from $T_{\rm C}$ = 154 K.

    It is also interesting to point out that the specific heat measurements \cite{Sekine2015} suggest that the rattling motion of the Eu atoms is not prominent in EuFe$_4$As$_{12}$.
  This has been generally explained in terms of the  ionic radius of Eu$^{2+}$ ions greater than those of  trivalent rare-earth ions, which gives rise to a less space for the rattling motion of the Eu$^{2+}$ ions in the Fe$_4$As$_{12}$ cage.
   However, as the cubic electric field at the rare-earth site is also considered to be one of the keys for the rattling motion, the lowering of the cubic symmetry at the Eu site in EuFe$_4$As$_{12}$ may result in restricting the rattling motion of the Eu atoms.     
   Even from this point of view,  as described above, further detailed studies on the crystal structure  at low temperatures  are requested,  which may provide further insights into the "rattling" physics in filled skutterudite compounds.

      At the end of this sub-subsection, we show the temperature dependence of $1/T_1$ measured at the central line of the $^{153}$Eu zero-field NMR spectrum.
      As shown in  Fig. \ref{fig:TEuNMR}(d),  $1/T_1$ is proportional to temperature, obeying a Korringa law $1/T_1T$ = 28 (sK)$^{-1}$ expected for a metallic state.             
    This result confirms the metallic ground state in the ferrimagnetic state from a microscopic point of view, consistent with the resistivity measurements \cite{Sekine2009,Sekine2015}.

 \subsection{$^{75}$As NQR  in the paramagnetic state} 

 \begin{figure}[tb]
\includegraphics[width=\columnwidth]{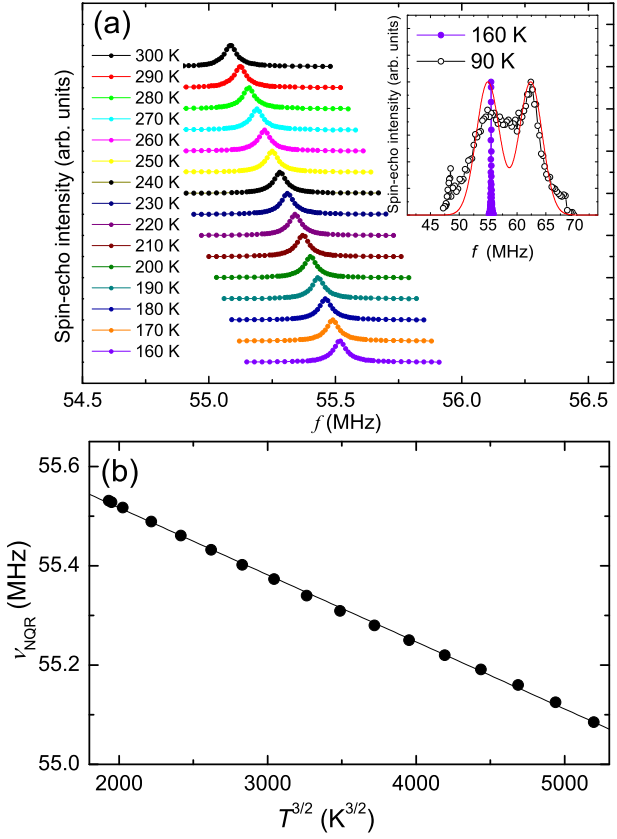} 
\caption{(a) Temperature dependence of the $^{75}$As-NQR spectrum for EuFe$_4$As$_{12}$ in the paramagnetic state. 
The inset shows a typical  $^{75}$As-NQR spectrum measured at $T$ = 90 K in the magnetically ordered state.
The red curve is a calculated result with $\nu_{\rm NQR}$ = 58. 7 MHz and $|$$B_{\rm int}$$|$ = 5.1 kOe at the As site.
(b) Temperature dependence of $^{75}$As-NQR frequency $\nu_{\rm NQR}$ in the paramagnetic state as a function of $T ^{3/2}$. The solid line is the fitting result (see text). 
}
\label{fig:NQR}
\end{figure}    

        Figure \ \ref{fig:NQR} (a) shows the temperature dependence of the $^{75}$As NQR spectrum from 160 K to 300 K in the paramagnetic state.
        In NQR spectrum under zero magnetic field for $^{75}$As nuclei with $I$ = 3/2, one expects a single transition line at a frequency of $\nu_{\rm NQR} = \nu_{\rm Q}\sqrt{1+{\eta}^2/3}$.   
         The observed lines are sharp with the nearly temperature independent line width ($FWHM$ $\sim$ 65 kHz), which is smaller than $FWHM$ $\sim$ 140 kHz and is greater than  $\sim$ 33 kHz for high quality samples of SrFe$_4$As$_{12}$ \cite{Ding2018} and LaFe$_4$As$_{12}$ \cite{Nowak2009}, respectively, but is much smaller than $\sim$ 400 kHz in SrOs$_4$As$_{12}$ having a lesser degree of homogeneity of crystals \cite{Ding2019}. 
      This indicates again the high quality of the samples and also the Eu filling factor close to unity.  
      As shown in the figure, the peak position slightly shifts to lower frequency with increasing temperature without showing any sudden changes, indicating that there is no any structural anomaly in the paramagnetic state of EuFe$_4$As$_{12}$.
      On the other hand, below $T_{\rm C}$, the spectrum becomes broader and the signal intensity decreases.
      Although we could not measure the spectrum around $T_{\rm C}$ because of the poor signal intensity due to the shortening of nuclear spin-spin relaxation time $T_2$ originating from the phase transition, we were able to  measure the spectrum at $T$ = 90 K.
   The very broad spectrum with the two-peak structure is observed as shown in the inset of Fig. \ \ref{fig:NQR}(a) where we also plot the spectrum ($T$ = 160 K) observed in the paramagnetic state for comparison. 
     One possible explanation of the two-peak structure is to introduce an internal field at the As site from the Eu and Fe ordered moments in the ferrimagnetic state. 
  It is also important to point out that  the center of mass of the spectrum shifts to higher frequency, indicative of an increase of $\nu_{\rm NQR}$.
    Based on this consideration, we have calculated $^{75}$As NQR spectrum assuming $|B_{\rm int}|$ = 5.1 kOe and $\nu_{\rm NQR}$ = 58.7 MHz with a line broadening of 4 MHz. 
      The red curve is the calculated spectrum, which seems to capture the characteristic shape of the spectrum. 
      Although we cannot determine the direction of $|B_{\rm int}|$ as well as its sign, the large change of $\nu_{\rm NQR}$ from 55.5 - 55.1 MHz in the paramagnetic state to 58.7 MHz at 90 K would be consistent with  the observation of the local distortion related to a possible  structural phase transition suggested by the $^{153}$Eu NMR measurements.
                     
       The temperature dependence of $\nu_{\rm NQR}$ determined from the peak positions of the NQR spectra is shown in Fig.\ \ref{fig:NQR} (b). 
      Similar temperature dependences of $\nu_{\rm NQR}$ have been observed in SrFe$_4$As$_{12}$ \cite{Ding2018}, SrOs$_4$As$_{12}$ \cite{Ding2019} and also in other filled skutterudite compounds \cite{Matsumura2007,Shimizu2007,Magishi2014,Nowak2009,Nowak2011,Yogi2014} where the temperature dependence is found to obey an empirical relation $\nu_{\rm NQR}(T) = \nu_{\rm NQR}(0)(1-\alpha_{\rm Q} T^{3/2})$ with a fitting parameter $\alpha_{\rm Q}$ \cite{alphaQ}.
     As shown by the curve in Fig.\ \ref{fig:NQR} (b), the temperature dependence of  $\nu_{\rm NQR}$ at As sites in EuFe$_4$As$_{12}$ also follows the relation with $\alpha_{\rm Q} = 2.24 \times 10^{-6}$  K$^{-3/2}$. 
    The value of $\alpha_{\rm Q} = 2.24 \times 10^{-6}$  K$^{-3/2}$ is similar to the values for SrFe$_4$As$_{12}$ (3.21 $\times 10^{-6}$  K$^{-3/2}$) \cite{Ding2018} and SrOs$_4$As$_{12}$ (2.09 $\times 10^{-6}$  K$^{-3/2}$) \cite{Ding2019}. 
    
    It is noted that one cannot determine the values of $\nu_{\rm{Q}}$ and $\eta$ for the As nuclei separately from only the NQR spectrum measurements.
    The values of $\eta$ = 0.4-0.45 have been reported in the isostructural compounds Sr$M_4$As$_{12}$ ($M$ = Fe, Os) \cite{Ding2018, Ding2019}.
    Since the crystal structure of EuFe$_4$As$_{12}$ in the paramagnetic state is the same with those compounds, we expect the similar value of $\eta$ in the present compound.

\begin{figure}[tb]
\includegraphics[width=\columnwidth]{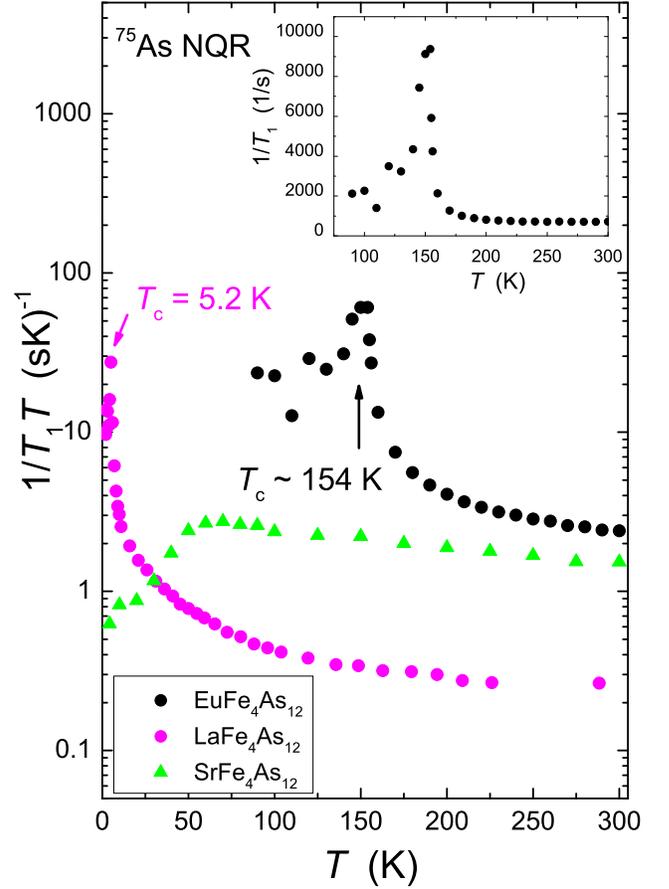} 
\caption{Temperature dependence of $^{75}$As 1/$T_1T$ in EuFe$_4$As$_{12}$ (black circles), together with those in LaFe$_4$As$_{12}$ (magenta circles) from Ref. \cite{Asaki2011} and SrFe$_4$As$_{12}$ (green triangles) from Ref. \cite{Ding2018}.  
   The inset shows the temperature dependence of 1/$T_1$. }   
\label{fig:T1}
\end{figure}

     To investigate the magnetic fluctuations in the paramagnetic state in EuFe$_4$As$_{12}$, we have measured the temperature dependence of  $^{75}$As spin-lattice relaxation rate 1/$T_1$ in zero magnetic field.
      Figure \ \ref{fig:T1} shows the temperature dependence of 1/$T_1T$.
      For comparison, we also plotted the 1/$T_1T$ data for the itinerant ferromagnet LaFe$_4$As$_{12}$ reported previously \cite{Asaki2011} where a clear peak in 1/$T_1T$ can be observed at a Curie temperature of 5.2 K.
         With decreasing temperature, 1/$T_1T$ for  EuFe$_4$As$_{12}$ increases gradually and shows a divergent behavior around 153 K, corresponding to the ferrimagnetic phase transition at $T_{\rm C} \sim$  154 K. 
        The values of 1/$T_1T$ for EuFe$_4$As$_{12}$ above $T_{\rm C}$ are almost one order of magnitude greater than those in LaFe$_4$As$_{12}$ at the temperature region. 
        Since there is no 4$f$ electrons in LaFe$_4$As$_{12}$, the largely enhanced $^{75}$As relaxation in EuFe$_4$As$_{12}$ could mainly originate from the fluctuations of the Eu 4$f$ electron moments.
    According to Moriya \cite{Moriya1956}, when the magnetic fluctuations are dominated by the paramagnetic fluctuations of local moments,  1/$T_1$ is expected to be a constant well above magnetic ordering temperature.
   As shown in the inset of  Fig. \ \ref{fig:T1}, 1/$T_1$ is nearly independent of temperature at high temperatures.
   This indicates that the magnetic fluctuations in the paramagnetic state of EuFe$_4$As$_{12}$ are characterized by the fluctuations of the local moment nature of the Eu 4$f$ electron spins. 
   This is consistent with the local moment picture of the Eu spins indicated by the temperature dependence of $M/H$ discussed above. 
           
         
         It is also interesting to compare the $1/T_1T$ data for LaFe$_4$As$_{12}$ with those for the paramagnetic metal SrFe$_4$As$_{12}$ (plotted by the green triangles in Fig.~\ \ref{fig:T1}) where the existence of ferromagnetic spin fluctuations was reported \cite{Ding2019}. 
         The values of 1/$T_1T$ of  LaFe$_4$As$_{12}$ well above $T_{\rm C}$ = 5.2 K  are much less than those in SrFe$_4$As$_{12}$.              
         Although we do not have the detailed information about the local density of states at the As sites for both compounds, it is interesting if such reduction of 1/$T_1T$ in LaFe$_4$As$_{12}$ could be attributed to the suppression of ferromagnetic spin fluctuations, resulting in the ferromagnetic order at $T_{\rm C}$ = 5.2 K.   
         Further studies, especially electronic structure calculations, are required to address this issue.

      \section{Summary}
     In summary, we performed $^{75}$As NQR and $^{153}$Eu NMR measurements on the filled skutterudite compound EuFe$_4$As$_{12}$ with $T_{\rm C}$ = 154 K. 
      We observed the $^{153}$Eu NMR spectrum in the magnetically ordered state, which reveals that the Eu$^{2+}$ ordered moments are close to 7 $\mu_{\rm B}$.
      From the external magnetic field dependence of $^{153}$Eu NMR spectrum observed in the magnetically ordered state, we found that the Eu ordered moments ferromagnetically align the magnetic field direction without any canting component. 
      Taking the magnetization data into consideration, we determined the magnetic ground state of EuFe$_4$As$_{12}$ to be ferrimagnetic in which the Fe 3$d$ moment and the Eu 4$f$ moment are magnetically ordered with antiferromagnetic coupling.      
    The observed $^{153}$Eu NMR spectrum shows quadrupole split lines which are not expected at the Eu site in the cubic structure ($Im\overline{3}$)  determined by the XRD measurements at room temperature, suggesting the lowering of the local symmetry at the Eu site at low temperatures.
    The temperature dependence of $^{75}$As 1/$T_1T$ suggests that the magnetic fluctuations in the paramagnetic state are dominated by the Eu 4$f$ electron spins which are well described by the local moment picture.
    It is shown that $^{153}$Eu NMR can be a unique tool in determining the magnetic structure in the Eu compound.
   It is interesting to study other Eu based magnetic compounds, such as EuFe$_4$Sb$_{12}$, EuOs$_4$Sb$_{12}$, EuFe$_4$P$_{12}$, and EuOs$_4$P$_{12}$ to gain deeper understanding about the interaction between the $d$ electrons and the Eu 4$f$ electrons.

   \section{Acknowledgments} 

The authors would like to acknowledge S. Pakhira and D. C.  Johnston for the determination of the chemical composition by EDX measurements at Ames laboratory.  The research was supported by the U.S. Department of Energy (DOE), Office of Basic Energy Sciences, Division of Materials Sciences and Engineering. Ames Laboratory is operated for the U.S. DOE by Iowa State University under Contract No.~DE-AC02-07CH11358.
    Part of this work was supported by JSPS KAKENHI Grant Number 23340092 and 19K03735.


\begin{thebibliography}{10}
\bibitem{Harima_Review} H. Sato, H. Sugawara, Y. Aoki, and H. Harima, Magnetic Properties of Filled Skutterudites in {\it Handbook of Magnetic Materials} edited by K.H.J. Buschow (The Netherlands: Elsevier, 2009), Vol 18, pp. 1 –- 110, and references there in.
\bibitem{Stewart1984} G. R. Stewart, Rev. Mod. Phys. {\bf 56}, 755 (1984).
\bibitem{Radousky} H. B. Radousky, {\it Magnetism in Heavy Fermion Systems} (World Scientific, Singapore, 2000).
\bibitem{Muller2001} K.-H. M\"uller and V. N. Narozhnyi. Rep. Prog. Phys. {\bf 64}, 943 (2001).
\bibitem{Sales2005} B. C. Sales, Filled skutterudites, in $Handbook$ $on$ $the$ $Physics$ $and$ $Chemistry$ $of$ $the$ $Rare$ $Earths$, edited by K. A. Gschneidner Jr., J.-C. B\"unzli, and V. K. Pecharsky (Elsevier Science, Amsterdam, 2003), Vol. 33, Chap. 211, and references there in.
\bibitem{Kirchner2020}S. Kirchner, S. Paschen, Q. Chen, S. Wirth, D. Feng, J. D. Thompson, and Q. Si, Rev. Mod. Phys. {\bf 92}, 011002 (2020).



\bibitem{Jeevan2008} H. S. Jeevan, Z. Hossain, D. Kasinathan, H. Rosner, C. Geibel, and P. Gegenwart,  Phys. Rev. B {\bf 78}, 052502 (2008).
\bibitem{Tereshima2009} T. Terashima, M. Kimata, H. Satsukawa, A. Harada, K. Hazama, S. Uji, H. S. Suzuki, T. Matsumoto, and K. Murata,  J. Phys. Soc. Jpn. {\bf 78}, 083701 (2009). 
\bibitem{Jin2016} W. T. Jin, Y. Xiao, Z. Bukowski, Y. Su, S. Nandi, A. P. Sazonov, M. Meven, O. Zaharko, S. Demirdis, K. Nemkovski, K. Schmalzl, L. M. Tran, Z. Guguchia, E. Feng, Z. Fu, and Th. Br\"uckel, Phys. Rev. B {\bf 94}, 184513 (2016).

\bibitem{Zapf2017} S. Zapf and M. Dressel, Rep. Prog. Phys. {\bf 80}, 016501 (2017).

\bibitem{Sekine2009} C. Sekine, K. Akahira, K. Ito, and T. Yagi, J. Phys. Soc. Jpn. {\bf 78}, 093707 (2009).
\bibitem{Sekine2015} C. Sekine, K. Ito, K. Akihara, Y. Kawamura, Y. Q. Chen, H. Gotou, and K. Matsuhira, J. Phys.: Conf. Ser. {\bf 592}, 012032 (2015). 
\bibitem{Sekine2015Sr} C. Sekine, T. Ishizaka, K. Nishine, Y. Kawamura, J. Hayashi, K. Takeda, H. Gotou, and Z. Hiroi, Phys. Procedia {\bf 75}, 383 (2015).
\bibitem{Nishine2017} K. Nishine, Y. Kawamura, J. Hayashi, and C. Sekine, Jpn. J. Appl. Phys. {\bf 56}, 05FB01 (2017).
\bibitem{Ding2018} Q.-P. Ding, K. Rana, K. Nishine, Y. Kawamura, J. Hayashi, C. Sekine, and Y. Furukawa,  Phys. Rev. B {\bf 98}, 155149 (2018).

\bibitem{Tatsuoka2008} S. Tatsuoka, H. Sato, K. Tanaka, M. Ueda, D. Kikuchi, H. Aoki, T. Ikeno, K. Kuwahara, Y. Aoki, H. Sugawara, and H. Harima, J. Phys. Soc. Jpn. {\bf 77}, 033701 (2008).

\bibitem{Namiki2010} T. Namiki, C. Sekine, K. Matsuhira, M. Wakeshima, and I. Shirotani,  J. Phys. Soc. Jpn. {\bf 79}, 074714 (2010).

\bibitem{Danebrock1996} M. E. Danebrock, C. B. H. Evers, and W. Jeitschko, J. Phys. Chem. Solids {\bf 57}, 381 (1996).
\bibitem{Takeda2000} N. Takeda and M. Ishikawa, J. Phys. Soc. Jpn.  {\bf 69}, 868 (2000).
\bibitem{Bauer2001} E. Bauer, St. Berger, A. Galatanu, M. Galli, H. Michor, G. Hilscher, Ch. Paul, B. Ni, M. M. Abd-Elmeguid, V. H. Tran, A. Grytsiv, and P. Rogl, Phys. Rev. B {\bf 63}, 224414 (2001).
\bibitem{Kihou2004} K. Kihou, I. Shirotani, Y. Shimaya, C. Sekine, and T. Yagi, Mater. Res. Bull. {\bf 39}, 317 (2004).
\bibitem{Bauer2004} E. D. Bauer, A. Slebarski, N. A. Frederick, W. M. Yuhasz, M. B. Maple, D. Cao, F. Bridges, G. Giester, and P. Rogl, J. Phys.: Condens. Matter {\bf 16}, 5095 (2004).
\bibitem{Gerard1983} A. Gerard, F. Grandjean, J. A. Hodges, D. J. Braun, and W. Jeitschko, J. Phys. C {\bf 16}, 2797 (1983).
\bibitem{Grandjean1984} F. Grandjean, A. G\'{e}rald, D. J. Braun, and W. Jeitschko, J. Phys. Chem. Solids {\bf 45}, 877 (1984).
\bibitem{Krishnamurthy2007} V. V. Krishnamurthy, J. C. Lang, D. Haskel, D. J. Keavney, G. Srajer, J. L. Robertson, B. C. Sales, D. G. Mandrus, D. J. Singh, and D. I. Bilc, Phys. Rev. Lett. {\bf 98}, 126403 (2007).
\bibitem{Slichterbook} C. P. Slichter, Principles~of~Magnetic~Resonance, 3rd ed. (Springer, New York, 1990).
\bibitem{Freeman1965} A. J. Freeman and R. E. Watson, in $Magnetism$, edited by G. T. Rado and H. Suhl (Academic Press, New York, 1965), Vol. IIA, Ch. 4, pp. 167--305.
\bibitem{Higa2017} N. Higa, Q.-P. Ding, M. Yogi, N. S. Sangeetha, M. Hedo, T. Nakama, Y. \=Onuki, D. C. Johnston, and Y. Furukawa, Phys. Rev. B {\bf 96}, 024405 (2017).
\bibitem{Ding2017} Q.-P. Ding, N. Higa, N. S. Sangeetha, D. C. Johnston, and Y. Furukawa, Phys. Rev. B {\bf 95}, 184404 (2017).
\bibitem{Yogi2013} M. Yogi, S. Nakamura, N. Higa, H. Niki, Y. Hirose, Y. \=Onuki, and H. Harima, J. Phys. Soc. Jpn. {\bf 82}, 103701 (2013).
\bibitem{Reehuis1992} M. Reehuis, W. Jeitschko, M. H. M\"oller, and P. J. Brown, J. Phys. Chem. Solids {\bf 53}, 687 (1992).
\bibitem{Tan2016} X. Tan, G. Fabbris, D. Haskel, A. A. Yaroslavtsev, H. Cao, C. M. Thompson, K. Kovnir, A. P. Menushenkov, R. V. Chernikov, V. O. Garlea, and M. Shatruk, J. Am. Chem. Soc. {\bf 138}, 2724 (2016).
\bibitem{VESTA} The crystal structure was drawn by using VESTA [K. Momma and F. Izumi, J. Appl. Crystallogr., {\bf 44}, 1272-1276 (2011)].

\bibitem{Niki2020} H. Niki, S. Nakamura, N. Higa, M. Yogi, A Nakamura, K. Niki, T.  Maehira, M. Hedo, T. Nakama, and Y.  \=Onuki, JPS Conf. Proc. {\bf 29}, 012007 (2020).
\bibitem{Gippius2006} A. Gippius,  M. Baenitz, E. Morozova, A. Leithe-Jasper, W. Schnelle, A. Shevelkov, E. Alkaev,  A. Rabis, J. A. Mydosh, Y. Grin, and F. Steglich, J. J. Magn. Magn. Mater. {\bf 300}, e403 (2006).
\bibitem{Magishi2006} K. Magishi, Y. Nakai, K. Ishida, H. Sugawara, I. Mori, T. Saito, and K. Koyama, J. Phys. Soc. Jpn. {\bf 75}, 023701 (2006).
\bibitem{Yamamoto2008} A. Yamamoto, S. Inemura, S. Wada, K. Ishida, I. Shirotani, and C. Sekine, J. Phys.: Condens. Matter  {\bf 20},  195214 (2008).

\bibitem{ShanlerEuFe4As12} A. Shankar, D. P. Rai, Sandeep, J. Maibam, and R. K. Thapa, AIP Conf. Proc. {\bf 1661}, 070010 (2015).
\bibitem{ShanlerEuFe4P12} A. Shankar and R. K. Thapa, Physica B {\bf 427}, 31–36 (2013).
\bibitem{ShanlerEuFe4Sb12} A.Shankar, D. P. Rai, Sandeep, and R. K. Thapa, Phys. Procedia {\bf 54}, 127 (2014). 


\bibitem{alphaQ} S. Takagi, H. Muraoka, T. D. Matsuda, Y. Haga, S. Kambe, R. E. Walstedt, E. Yamamoto, and \=Onuki, J. Phys. Soc. Jpn. {\bf 73},  469 (2004).
\bibitem{Nowak2009}  B. Nowak, O. {\.Z}oga\l, A. Pietraszko, R. E. Baumbach, M. B. Maple, and Z. Henkie, Phys. Rev. B {\bf 79}, 214411 (2009).

\bibitem{PrRu4P12_0} C. H. Lee, H. Matsuhata, A. Yamamoto, T. Ohta, H. Takazawa, K. Ueno, C. Sekine, I. Shirotani, and T. Hirayama, J. Phys.: Condens. Matter {\bf 13}, L45 (2001).
\bibitem{PrRu4P12_1} C. H. Lee, H. Matsuhata, H. Yamaguchi, C. Sekine, K. Kihou, T. Suzuki, T. Noro, and I. Shirotani, Phys. Rev. B {\bf 70}, 153105 (2004).
\bibitem{PrRu4P12_2} C. H. Lee, H. Matsuhata, H. Yamaguchi, C. Sekine, K. Kihou, and I. Shirotani, J. Mag. Mag. Mat., {\bf 272-276}, 426 (2004). 


\bibitem{Ding2019} Q.-P. Ding, K. Nishine, Y. Kawamura, J. Hayashi, C. Sekine, and Y. Furukawa,  Phys. Rev. B {\bf 100}, 054516 (2019).

\bibitem{Matsumura2007} M. Matsumura, G. Hyoudou, M. Itoh, H. Kato, T. Nishioka, E. Matsuoka, H. Tou, T. Takabatake, and M. Sera, J. Phys. Soc. Jpn. {\bf 76}, 084716 (2007).
\bibitem{Shimizu2007} M. Shimizu, H. Amanuma, K. Hachitani, H. Fukazawa, Y. Kohori, T. Namiki, C. Sekine, and I. Shirotani, J. Phys. Soc. Jpn. {\bf 76}, 104705 (2007).
\bibitem{Magishi2014} K. Magishi, R. Watanabe, A. Hisada, T. Saito, K. Koyama, T. Saito, R. Higashinaka, Y. Aoki, and H. Sato, J. Phys. Soc. Jpn. {\bf 83}, 084712 (2014).
\bibitem{Nowak2011} B. Nowak, O. {\.Z}oga\l, Z. Henkie, and M.B. Maple, Solid State Commun.  {\bf 151}, 550 (2011).
\bibitem{Yogi2014} M. Yogi, H. Niki, T. Kawata, and C. Sekine, JPS Conf. Proc. {\bf 3},  011046 (2014).
\bibitem{Asaki2011} K. Asaki, H. Kotegawa, H. Tou, S. Tatsuoka, R. Higashinaka, T. Namiki, and H. Sato, J. Phys. Soc. Jpn. {\bf 80}, SA033 (2011).
\bibitem{Moriya1956} T. Moriya, Prog. Theor. Phys. {\bf 16}, 23 (1956).




\end{thebibliography}
\end{document}